\documentclass{article}

\usepackage{lineno}
\usepackage{amsmath}  
\usepackage{etoolbox} 

\usepackage{amsfonts}
\usepackage{amssymb}
\usepackage{amsthm}
\usepackage{graphicx}
\usepackage[hidelinks]{hyperref}
\usepackage{subcaption}
\usepackage{tikz}
\usepackage{algpseudocode}
\usepackage{algorithm}
\usepackage{enumitem}
\usepackage{natbib}
\setcitestyle{authoryear, open={(},close={)}}
\usepackage{orcidlink}

\newtheorem{theorem}{Theorem}
\newtheorem{corollary}{Corollary}

\newcommand{\np}{\textit{NP}}
\newcommand{\nni}{\mathrm{NNI}}
\newcommand{\spr}{\mathrm{SPR}}
\newcommand{\tbr}{\mathrm{TBR}}
\newcommand{\rf}{\mathrm{RF}}

\title{Tree rearrangement graphs admit paths of decreasing Robinson-Foulds distance}

\author{
    Lena~Collienne$^{1,*}$\,\orcidlink{0000-0003-4215-8263},
    Frederick A Matsen IV$^{1,2,3,4}$\,\orcidlink{0000-0003-0607-6025}\\[8pt]
    \parbox{\linewidth}{\centering
    \textit{$^{1}$~Computational Biology Program, Fred Hutchinson Cancer Research Center, Seattle, Washington, USA;\\
    $^{2}$~Howard Hughes Medical Institute, Fred Hutchinson Cancer Research Center, Seattle, Washington, USA;\\
    $^{3}$~Department of Statistics, University of Washington, Seattle, USA;\\
    $^{4}$~Department of Genome Sciences, University of Washington, Seattle, USA;\\
    *Correspondence: lena@lenacoll.de}}}

\begin{document}

\maketitle

\begin{abstract}
    Tree rearrangements such as Nearest Neighbor Interchange (NNI) and Subtree Prune and Regraft (SPR) are commonly used to explore phylogenetic treespace.
    Computing distances based on them, however, is often intractable, so the efficiently computable Robinson-Foulds (RF) distance is used in practice.
    We investigate how the RF distance behaves along paths in the NNI and SPR graphs, where trees are nodes and edges represent single rearrangements.
    We show that any two trees are connected by a path along which the RF distance to the target decreases monotonically in the NNI graph and strictly in the SPR graph; we also show that a strictly decreasing NNI path does not always exist.
\end{abstract}


\section{Introduction}

Phylogenetic trees are complex objects that can be viewed from many perspectives~\citep{st_john_review_2017}.
One perspective considers a tree as a collection of bipartitions: statements about which leaves are separated from other leaves.
These bipartitions are called splits; trees are in one-to-one correspondence with their sets of splits.
One can turn this perspective into a distance between phylogenetic trees by considering the size of the symmetric difference between their split collections.
This distance, formally defined below, is called the Robinson-Foulds ($\rf$) distance.

Another perspective on phylogenetic trees is as an organized collection of subtrees.
A natural operation on a tree from this perspective is to separate a subtree from one location and move it to another location in the tree.
If that movement just swaps subtrees across an internal edge, this is called a Nearest Neighbor Interchange ($\nni$).
If the movement reattaches the subtree on an arbitrary edge, this is called a Subtree Prune and Regraft ($\spr$).
If the movement allows the reattachment point of the pruned subtree itself to be arbitrary as well, this is called Tree Bisection and Reconnection ($\tbr$).
One can build a graph such that the vertices of the graph are phylogenetic trees, and adjacency in the graph is determined by whether one tree can be transformed to another by one of these operations.

The corresponding graph distance for the $\nni$ (resp. $\spr$) adjacency is called the $\nni$ (resp. $\spr$) distance.
For both $\nni$ and $\spr$ it has been shown that computing the distance between two trees is $\np$-hard~\citep{dasgupta_computing_2000,bordewich_computational_2005,hickey_spr_2008}.
In contrast, the Robinson-Foulds ($\rf$) distance is computable in time linear in the number of leaves~\citep{day_optimal_1985}.

The intractable rearrangement-based distances are generally considered preferable to the RF distance.
First, these distances have biological interpretations in terms of hybridizations~\citep{bordewich_computational_2005,bordewich_computing_2007}.
Second, these distances correspond to the optimization steps of popular maximum likelihood methods and proposal moves for Markov chain Monte Carlo (MCMC) Bayesian phylogenetic algorithms.
However, as stated above, these distances are hard to compute.

Consequently, exploration of large collections of trees \cite[e.g.][analyzing MCMC chains]{gao_biological_2025} often uses both RF and SPR distances.
RF can be computed for all pairs in these collections.
SPR is intractable for large datasets, so researchers resort to approximations and subsampling.
The conclusions drawn from RF and SPR distances often differ substantially~\citep{brusselmans_importance_2024,gao_biological_2025}.

Hence, we are motivated to understand the relationship between $\rf$ and tree rearrangement distances.
Previous work has shown that the RF distance is not robust with respect to the $\spr$ rearrangement in the sense that trees with a small $\spr$ distance can be very distant under Robinson-Foulds.
For example, a single $\spr$ move can connect two trees with maximum $\rf$ distance, if the pruned subtree contains just one leaf that gets reattached on the opposite end of a tree~\citep{bocker_generalized_2013}.
On the other hand, because $\nni$ moves are more local in a tree and only change one edge, the $\nni$ distance between two trees is at most twice as large as their $\rf$ distance~\citep{steel_phylogeny_2016}.
The combinatorial structure of the neighborhood of a tree under these rearrangements, and how it relates to splits, has also been studied directly: \citet{bryant_splits_2004} characterized exactly which splits appear in the neighborhood of a tree under each of the $\rf$, $\nni$, $\spr$, and $\tbr$ metrics, and \citet{de_jong_neighborhoods_2016} gave exact and asymptotic counts of the trees at a given $\nni$, $\spr$, or $\rf$ distance from a fixed tree.

Despite these advances, our knowledge of the connection between these distances remains incomplete.
As described above, the NNI (resp. SPR) graph is the graph formed with trees as nodes and edges representing NNI (resp. SPR) moves.
The work described in the previous paragraph compares shortest paths in these graphs to the RF distance, and shows that these can be in conflict.
However, it does not address the question of how the RF distance changes on arbitrary paths through these graphs.

Specifically, given a pair of trees $T$ and $R$, is there a path between them in the $\nni$ or $\spr$ graph such that the RF distance between the trees along this path and $R$ is monotonically decreasing?
A negative answer would argue that these measures are fundamentally at odds.
This question was posed as an open problem by Nick Goldman at the ``Algorithmic Advances and Implementation Challenges: Developing Practical Tools for Phylogenetic Inference'' workshop at the Institute for Computational and Experimental Research in Mathematics in Providence, RI in November 2024.

In this paper, we answer this question in the affirmative.
Specifically, we show that any two trees are connected by a path in the $\nni$ graph on which the RF distance to the target tree is monotonically decreasing along the path; we also exhibit a pair of trees for which it cannot be strictly decreasing.
In the $\spr$ graph, however, we can always find a path with strictly decreasing RF distance to the final tree.

\section{Definitions}

A \emph{phylogenetic tree} is a tree with leaves bijectively labeled by elements of a set $X$.
For simplicity, we refer to phylogenetic trees as \emph{trees}.
Unless otherwise stated, we assume that trees are unrooted and binary, i.e. all internal nodes are of degree three.

A \emph{split} $A|B$ of a tree $T$ with leaf labels in $X$ is a bipartition of $X$ into sets $A, B \neq \emptyset$ such that cutting an edge $e$ in $T$ results in two connected components with respective leaf sets $A$ and $B$.
We also say that $e$ \emph{induces} the split $A|B$.
The \emph{split set} $\Gamma(T)$ is the set of all splits of $T$, and it uniquely identifies $T$ \citep{robinson_comparison_1981}.
We call two phylogenetic trees $T$ and $R$ equal ($T = R$) if they represent the same set of splits.
We define the \emph{Robinson-Foulds} ($\rf$) distance between two trees $T$ and $R$ as the size of the symmetric difference of their split sets: $d_{\rf}(T,R) = |\Gamma(T) \Delta \Gamma(R)|$.
The $\rf$ distance for binary trees is sometimes defined as half of this quantity, i.e. $\frac{1}{2}|\Gamma(T) \Delta \Gamma(R)|$~\cite{bryant_splits_2004}, but we use the full symmetric difference here.
A split $A|B$ is called \emph{trivial} if either $A$ or $B$ contains only one element, otherwise it is called \emph{non-trivial}.
For a subset $A \subseteq X$, the \emph{subtree of $T$ induced by $A$} is the smallest connected subgraph of $T$ containing all leaves in $A$, with degree-$2$ vertices suppressed.

In this paper we focus on two tree rearrangement operations: Nearest Neighbor Interchange and Subtree Prune and Regraft.
Trees $T$ and $R$ are connected by a \emph{Nearest Neighbor Interchange} ($\nni$) move if there is an edge $e$ in $T$ and an edge $f$ in $R$ so that shrinking $e$ and $f$ to a single node results in equal (non-binary) trees.
One can also interpret an $\nni$ move as swapping two subtrees that are adjacent to opposite ends of the same internal edge.
A \emph{Subtree Prune and Regraft} ($\spr$) move on a tree $T$ begins by removing an edge $e$ to receive two connected components $T_1$ and $T_2$.
Then a new node $v$ is added on an edge $f$ in $T_2$.
If $T_1$ consists of a single leaf only, a new edge connects this leaf and $v$, otherwise a new edge connects the degree-$2$ node in $T_1$ and $v$, and any remaining degree-$2$ nodes in the resulting graph get suppressed.
Every $\nni$ move is also an $\spr$ move.

Tree rearrangements define a graph where nodes represent trees on $n$ leaves that are connected by an edge if the corresponding trees are connected by a tree rearrangement.
We call these graphs the $\nni$ graph and the $\spr$ graph.
A \emph{path} between trees $T$ and $R$ is a sequence $[T = T_0, T_1, \dots, T_{k-1}, T_k = R]$ of trees so that $T_i$ and $T_{i+1}$ are connected by a tree rearrangement.
The \emph{distance} between trees $T$ and $R$ in the $\nni$ or $\spr$ graph is the length of a shortest path connecting $T$ and $R$, or in other words, the minimum number of tree rearrangements required to transform one tree into the other.

\section{Robinson-Foulds distance along tree rearrangement paths}

In this section we show that given a source and a target tree, we can find paths in the $\nni$ graph along which the Robinson-Foulds distance to the target tree decreases monotonically.
For the $\spr$ graph we can even show the existence of paths with strictly decreasing RF distance.
\subsection*{Paths in the $\nni$ graph}

In the following, we show there is a path between any two trees $T$ and $R$ in the $\nni$ graph so that the $\rf$ distance to $R$ decreases along this path from $T$ to $R$.

\begin{theorem}
    For binary trees $T$ and $R$ on the same leaf set $X$ there is a path $[T_0 = T, T_1, \dots, T_k = R]$ in the $\nni$ graph with $d_{\rf}(T_{i+1}, R) \leq d_{\rf}(T_i, R)$ for all $i = 0, \dots, k-1$.
    \label{thm:NNI_decreasing_rf}
\end{theorem}

\begin{proof}
    We prove this theorem by induction on the number of leaves of $T$ and $R$.
    For the induction basis we consider trees $T$ and $R$ with four leaves and therefore one internal edge.
    With just one internal edge, the RF distance is either zero or two.
    If $d_{\rf}(T, R) = 0$, $T$ and $R$ are equal and the theorem is true with the path consisting of just one tree $T = R$.
    If $d_{\rf}(T, R) = 2$, $T$ and $R$ are not equal.
    Because all trees on four leaves are connected by an $\nni$ move, the path $[T_0 = T, T_1 = R]$ fulfills the conditions of the theorem.

    For the induction step we assume that the theorem is true for all pairs of trees with less than or equal to $n$ leaves.
    Let $T$ and $R$ be trees on $n+1$ leaves.
    We distinguish the case (i) that $T$ and $R$ share at least one non-trivial split from the case (ii) that they have no splits in common, i.e. they have maximum RF distance.

    \begin{enumerate}[label=(\roman*), leftmargin=0pt, align=left]
        \item Let $T$ and $R$ be trees that both have an internal edge inducing the same non-trivial split $A|B$.

            Let $T_a$ and $R_a$ be the trees that result from replacing the subtree induced by $A$ in $T$ and $R$ with a single leaf $a$.
            By the induction hypothesis, we can find an $\nni$ path from $T_a$ to $R_a$ on which the $\rf$ distance to $R_a$ does not increase.
            By performing the $\nni$ moves of the path from $T_a$ to $R_a$ on $T$, we receive a path from $T$ to a tree $T'$.
            Since these $\nni$ moves operate within the subtrees induced by $B$, all splits in the subtree induced by $A$ in $T$ as well as the split $A|B$ are preserved along this path.
            Let $\Gamma_A^T$ and $\Gamma_A^R$ be the splits induced by the subtrees of $T$ and $R$ induced by $A$.
            For the trees $T^j$ along the path from $T$ to $T'$, we have
            \begin{equation*}
                d_{\rf}(T^j, R) = |\Gamma_A^T \mathbin{\Delta} \Gamma_A^R| + d_{\rf}(T^j_a, R_a).
            \end{equation*}

            The first term is constant, and the second is non-increasing by the induction hypothesis, so the $\rf$ distance to $R$ is non-increasing along this path.
            As all splits in the subtree induced by $A$ in $T$ are preserved in $T'$, we can replace the subtree induced by $B$ in $T'$ and $R$ by a single leaf $b$, giving us trees $T'_b$ and $R_b$.
            Using the same argument as above, we can generate a path from $T'$ to $R$ along which the $\rf$ distance does not increase.
            By appending this path from $T'$ to $R$ to the path from $T$ to $T'$, we receive an $\nni$ path with non-increasing RF distance from $T$ to $R$.

        \item Let $T$ and $R$ be trees that share no non-trivial split.
            On any path from $T$ to $R$ in the $\nni$ graph there must be a tree that shares a split with $R$.
            Let $T'$ be the first tree on a path from $T$ to $R$ with this property.
            Since $T$ and $R$ share no non-trivial split, $d_{\rf}(T,R)$ is already the maximum possible RF distance for trees on $n+1$ leaves, so the RF distance to $R$ cannot increase along the trees preceding $T'$ on this path.
            Applying case (i) to $T'$ and $R$ gives us a path with the properties stated in the theorem.
    \end{enumerate}
\end{proof}

It is important to allow adjacent trees on the $\nni$ path to have the same $\rf$ distance to $R$ in Theorem~\ref{thm:NNI_decreasing_rf}.
We show in Theorem~\ref{thm:NNI_strictly_decreasing_rf} that we cannot assume strictly decreasing $\rf$ distances along an $\nni$ path.

\begin{theorem}
    There are binary trees $T$ and $R$ on the same leaf set $X$ for which no path $[T = T_0, T_1, \dots, T_k = R]$ exists in the $\nni$ graph with $d_{\rf}(T_{i+1}, R) < d_{\rf}(T_i, R)$.
    \label{thm:NNI_strictly_decreasing_rf}
\end{theorem}

Theorem~\ref{thm:NNI_strictly_decreasing_rf} is related to the observation of~\cite{bryant_splits_2004} that $d_{\rf}(T,R) \leq 2d_{\nni}(T,R)$, which we will use in the proof of this theorem.

\begin{proof}
    We assume to the contrary that for all trees $T$ and $R$ a path with strictly decreasing $\rf$ distance exists.
    Let $[T=T_0, T_1, \dots, T_k=R]$ be such a path.
    The $\rf$ distance between $T_i$ and $R$ is at least $2 (k-i)$, as the $\rf$ distance only takes even values.
    Then $d_{\rf}(T,R) \geq 2k$ for trees $T$ and $R$ that are connected by an $\nni$ path of length $k$.
    We also know that $d_{\nni}(T,R) \leq k$.
    Therefore, $d_{\rf}(T,R) \geq 2k \geq 2d_{\nni}(T,R)$.
    By \cite{bryant_splits_2004}, it is $d_{\rf}(T,R) \leq 2d_{\nni}(T,R)$ for all trees $T$ and $R$.
    This implies $2d_{\nni}(T,R) = d_{\rf}(T,R)$, which is not true~\cite{dasgupta_computing_2000}, leading us to a contradiction.

    Therefore, there must be a pair of trees $T$, $R$ that are not connected by a path of strictly decreasing $\rf$ distance in the $\nni$ graph.
\end{proof}

\section*{Paths in the $\spr$ graph}

$\spr$ moves are a generalization of $\nni$ moves, which means that every $\nni$ move is an $\spr$ move~\citep{allen_subtree_2001}.
By Theorem~\ref{thm:NNI_decreasing_rf} there is always a path in the $\spr$ graph on which the Robinson-Foulds distance to the final tree decreases monotonically.
In this section we show an even stronger result: There is a path between any two trees in the $\spr$ graph along which the RF distance to the destination tree is strictly decreasing.

\begin{theorem}
    For binary trees $T$ and $R$ on the same leaf set $X$ there is a path $[T = T_0, T_1, \dots, T_k = R]$ in the $\spr$ graph with $d_{\rf}(T_{i+1}, R) < d_{\rf}(T_i, R)$ for all $i = 0, \dots, k-1$.
    \label{thm:SPR_strictly_decreasing_rf}
\end{theorem}

The proof of Theorem~\ref{thm:SPR_strictly_decreasing_rf} that we present here implies an algorithm to compute paths in the $\spr$ graph of decreasing $\rf$ distance.

\begin{proof}
    We prove this theorem by induction on $d_{\rf}(T,R)$.
    For the induction basis we consider trees $T$ and $R$ with $d_{\rf}(T,R) = 0$.
    Then $T$ and $R$ are equal and the theorem is true with the path consisting of just one tree $T = R$.

    For the induction step we assume that the theorem is true for all pairs of trees with $\rf$ distance less than or equal to $m$.
    Let $T$ and $R$ be trees with $d_{\rf}(T,R) = m+1$.
    Then there is at least one split in $R$ that is not present in $T$.
    We define the size of a split $A|B$ to be $\min(|A|, |B|)$.
    Let $A|B$ be a non-trivial split of minimum size in $R$ that is not a split in $T$ and let $|A| \leq |B|$.
    Since $R$ is binary and $A|B$ is a non-trivial split in $R$, there are sets $A_1$ and $A_2$ with $A_1 \cup A_2 = A$ so that $A_1 | (A_2 \cup B)$ and $A_2 | (A_1 \cup B)$ are splits in $R$.
    By the minimality assumption on $A|B$, these splits are also present in $T$.

    \begin{figure}[h]
        \centering
        \includegraphics[width=\textwidth]{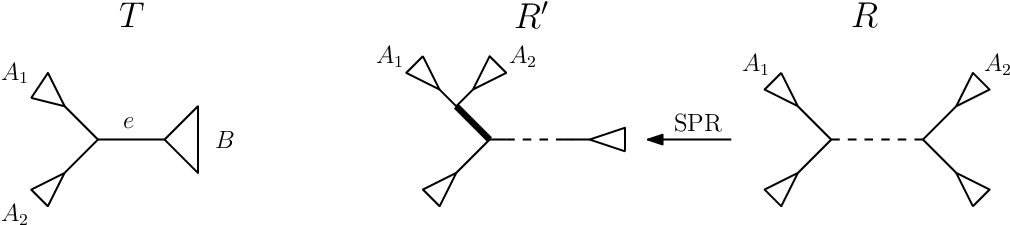}
        \caption{%
        Trees $R$, $T$, and $T'$ in the proof of Theorem~\ref{thm:SPR_strictly_decreasing_rf}.
        In $R$ on the left, edge $e$ induces the split $A|B$.
        Tree $T$ on the right and $T'$ in the middle are connected by an $\spr$ move on $T$ that moves the subtree induced by $A_2$.
        The dashed line in $T$ and $T'$ represents a section of the tree that might contain more leaves and edges.
        The dashed path in $T$ contains the edges inducing the splits $S$ changed by the $\spr$ move, none of which are present in $R$.
        The edge highlighted in bold in $T'$ induces the split $A|B$.}
        \label{fig:SPR_strictly_decreasing_rf}
    \end{figure}

    Let $T_{A_1}$ be the subtree of $T$ induced by $A_1$, and $T_{A_2}$ the subtree induced by $A_2$.
    We can perform an $\spr$ move on $T$ that prunes $T_{A_2}$ and reattaches it on the edge connecting $T_{A_1}$ with the rest of the tree.
    This creates an edge that induces the split $A|B$.
    Let $T'$ be the tree resulting from this $\spr$ move.

    We now consider the differences in the split sets of $T$ and $T'$ when comparing to $R$.
    By our assumption that $A|B$ is not a split in $T$, there is a path of edges between the subtrees $T_{A_1}$ and $T_{A_2}$ in $T$ where each edge induces a split of the form $\left( A_1 \cup C \right) | \left( A_2 \cup D \right)$ for some $C, D \neq \emptyset$.
    These are the only splits that change between $T$ and $T'$ when moving the subtree $T_{A_2}$, and we denote the set of these splits by $S$.

    Since $A|B = \left(A_1 \cup A_2\right)|B$ is a split in $R$, and $A|B$ is incompatible with all splits in $S$, no splits of $S$ are present in $R$.
    It follows that $S$ is a subset of the splits of $T$, but not of $R$.
    Therefore, $S$ is a subset of the symmetric difference of the split sets of $T$ and $R$.
    This means that changing only edges inducing splits in $S$, as is done between $T$ and $T'$, cannot increase the RF distance to $R$.
    Since the $\spr$ move between $T$ and $T'$ changes only splits in $S$ --- none of which are present in $R$ --- and creates the split $A|B$, which is present in $R$, the tree $T'$ has at least one more split in common with $R$ than $T$ does.
    Therefore, $d_{\rf}(T', R) < d_{\rf}(T,R) = m+1$, i.e. $d_{\rf}(T', R) \leq m$.

    Applying the induction hypothesis to $T'$ and $R$ yields a path $[T' = T_0', T_1', \dots, T_k' = R]$ in the $\spr$ graph with $d_{\rf}(T_{i+1}', R) < d_{\rf}(T_i', R)$ for all $i$.
    Since $T$ and $T' = T_0'$ are connected by an $\spr$ move with $d_{\rf}(T',R) < d_{\rf}(T,R)$, prepending $T$ gives a path $[T, T_0', \dots, T_k' = R]$ from $T$ to $R$ in the $\spr$ graph with strictly decreasing $\rf$ distance to $R$ throughout, completing the induction.
\end{proof}

Theorem~\ref{thm:SPR_strictly_decreasing_rf} can be read as a path-monotonic counterpart to the static characterization of the splits in the neighborhood of a tree given by~\citet{bryant_splits_2004}.
The splits altered by the single $\spr$ move in the proof, collected in the set $S$, are exactly the splits of $T$ that conflict with the target split $A|B$.
Lemma~3.1 of \citet{bryant_splits_2004} shows that the edges of a tree conflicting with a given split always form a connected subgraph; here that subgraph is precisely the path along which we reroute the subtree $T_{A_2}$.
Because every split in $S$ conflicts with $A|B \in \Gamma(R)$, none of them is a split of $R$, which is precisely what allows the move to strictly decrease the $\rf$ distance to $R$.
Bryant's characterization is static, describing which splits lie within a fixed distance of a tree; our result is dynamic, showing how to traverse the neighborhood so that agreement with the target tree only increases.

Tree Bisection and Reconnection ($\tbr$) is another tree rearrangement operation and generalizes $\spr$.
It starts by removing an edge and suppressing any resulting degree-$2$ nodes, and we receive two connected components.
In each connected component we choose a reconnection node.
If the component consists of a single leaf only, this is the reconnection node, otherwise we introduce a new node on an arbitrary edge.
By adding an edge between the two reconnection nodes, we finish the $\tbr$ move and receive a binary phylogenetic tree.
As all $\spr$ moves are $\tbr$ moves~\citep{allen_subtree_2001}, we get the following corollary.

\begin{corollary}
    For arbitrary trees $T$ and $R$ there is a path $[T = T_0, T_1, \dots, T_k = R]$ in the $\tbr$ graph with $d_{\rf}(T_{i+1}, R) < d_{\rf}(T_i, R)$ for all $i = 0, \dots, k-1$.
    \label{cor:TBR_strictly_decreasing_rf}
\end{corollary}

\section{Discussion}

In this paper we show that for any two trees $T$ and $R$ there is a path in the $\nni$ graph along which the Robinson-Foulds distance to $R$ decreases monotonically.
We also show that we cannot guarantee that the $\rf$ distance will strictly decrease along an $\nni$ path.
In the $\spr$ graph, however, there is a path with strictly decreasing RF distance to the destination tree.
These results give some insight into the relationship between Robinson-Foulds and tree rearrangement-based distances.

The constructive proof of Theorem~\ref{thm:SPR_strictly_decreasing_rf} can be used to generate paths along which the $\rf$ distance to the destination tree decreases.
These paths are generally not shortest paths in the $\spr$ graph, but they do preserve all shared splits.
While it is known that shortest $\spr$ paths may break shared splits~\citep{whidden_calculating_2019}, paths of decreasing $\rf$ distance preserve splits.

Our results also bear on the common practice of exploring tree space via randomized tree rearrangements, whether as local search moves in heuristic maximum likelihood search or as proposal moves in MCMC for Bayesian phylogenetics~\citep{whidden_quantifying_2015,gao_biological_2025}.
In such settings the true rearrangement distance to a reference tree is intractable to compute at each step, so the $\rf$ distance is a natural, efficiently computable proxy for guiding a search.
Theorems~\ref{thm:NNI_decreasing_rf} and~\ref{thm:SPR_strictly_decreasing_rf} show that such $\rf$-guided search is not fundamentally misguided: a path of non-increasing (or, for $\spr$, strictly decreasing) $\rf$ distance to the target always exists, even though following it greedily need not yield a shortest rearrangement path.

In future research we aim to investigate how good an approximation of the $\spr$ distance paths of decreasing $\rf$ distance provide.
The proof of Theorem~\ref{thm:SPR_strictly_decreasing_rf} implies an algorithm to construct those paths, but there are many paths with this property.
How to compute a path with decreasing $\rf$ distance that also minimizes the $\spr$ (or $\tbr$) distance remains an open problem.
Another future direction is to study paths of decreasing $\rf$ distance in rooted $\spr$, or in related rearrangement operations on ranked time trees such as the ranked $\spr$ of~\citet{collienne_ranked_2024}.
Under rooted $\spr$, common splits are preserved along shortest paths, suggesting that approximation algorithms using decreasing $\rf$ distances to the target tree could be effective in this setting.

\section{Data Availability}

Data sharing is not applicable to this article as no new data were created or analyzed in this study.

\section{Acknowledgments}

We would like to thank Nick Goldman for stating the questions of whether the Robinson-Foulds distance (strictly) decreases along paths in the $\nni$ and $\spr$ graphs, for describing the search heuristic that motivated these questions, and Minh Bui for presenting it to us.

This work was supported by NIH grant R01-AI162611.
Frederick Matsen is an investigator of the Howard Hughes Medical Institute.
This material is based upon work supported by the National Science Foundation under Grant No. DMS-1929284 while the authors were in residence at the Institute for Computational and Experimental Research in Mathematics in Providence, RI, during the ``Algorithmic Advances and Implementation Challenges: Developing Practical Tools for Phylogenetic Inference'' program.

\bibliographystyle{abbrvnat}
\bibliography{main}

@inbook{dasgupta_computing_2000,
	series = {{DIMACS} {Series} in {Discrete} {Mathematics} and                         {Theoretical} {Computer} {Science}},
	title = {On {Computing} the {Nearest} {Neighbor} {Interchange} {Distance}},
	volume = {55},
	url = {http://www.ams.org/dimacs/055},
	doi = {10.1090/dimacs/055},
	publisher = {American Mathematical Society},
	language = {en},
	urldate = {2023-02-14},
	author = {DasGupta, Bhaskar and He, Xin and Jiang, Tao and Li, Ming and Tromp, John and Zhang, Louxin},
	month = dec,
	year = {2000},
	pages = {125--143},
}

@article{whidden_quantifying_2015,
	title = {Quantifying {MCMC} {Exploration} of {Phylogenetic} {Tree} {Space}},
	volume = {64},
	issn = {1063-5157},
	url = {https://doi.org/10.1093/sysbio/syv006},
	doi = {10.1093/sysbio/syv006},
	number = {3},
	urldate = {2023-01-30},
	journal = {Systematic Biology},
	author = {Whidden, Chris and Matsen, IV, Frederick A.},
	month = may,
	year = {2015},
	pages = {472--491},
}

@article{hickey_spr_2008,
	title = {{SPR} {Distance} {Computation} for {Unrooted} {Trees}},
	volume = {4},
	issn = {1176-9343},
	url = {https://doi.org/10.4137/EBO.S419},
	doi = {10.4137/EBO.S419},
	language = {en},
	urldate = {2021-05-12},
	journal = {Evolutionary Bioinformatics},
	publisher = {SAGE Publications Ltd STM},
	author = {Hickey, Glenn and Dehne, Frank and Rau-Chaplin, Andrew and Blouin, Christian},
	month = jan,
	year = {2008},
	pages = {EBO.S419},
}

@article{allen_subtree_2001,
	title = {Subtree {Transfer} {Operations} and {Their} {Induced} {Metrics} on {Evolutionary} {Trees}},
	volume = {5},
	issn = {0218-0006, 0219-3094},
	url = {http://link.springer.com/10.1007/s00026-001-8006-8},
	doi = {10.1007/s00026-001-8006-8},
	language = {en},
	number = {1},
	urldate = {2021-05-12},
	journal = {Annals of Combinatorics},
	author = {Allen, Benjamin L. and Steel, Mike},
	month = jun,
	year = {2001},
	pages = {1--15},
}

@article{whidden_calculating_2019,
	title = {Calculating the {Unrooted} {Subtree} {Prune}-and-{Regraft} {Distance}},
	volume = {16},
	issn = {1557-9964},
	doi = {10.1109/TCBB.2018.2802911},
	number = {3},
	journal = {IEEE/ACM Transactions on Computational Biology and Bioinformatics},
	author = {Whidden, Chris and Matsen, Frederick A.},
	month = may,
	year = {2019},
	note = {Conference Name: IEEE/ACM Transactions on Computational Biology and Bioinformatics},
	pages = {898--911},
}

@article{bordewich_computational_2005,
	title = {On the {Computational} {Complexity} of the {Rooted} {Subtree} {Prune} and {Regraft} {Distance}},
	volume = {8},
	issn = {0218-0006, 0219-3094},
	url = {http://link.springer.com/10.1007/s00026-004-0229-z},
	doi = {10.1007/s00026-004-0229-z},
	language = {en},
	number = {4},
	urldate = {2021-04-22},
	journal = {Annals of Combinatorics},
	author = {Bordewich, Magnus and Semple, Charles},
	month = jan,
	year = {2005},
	pages = {409--423},
}

@article{robinson_comparison_1981,
	title = {Comparison of {Phylogenetic} {Trees}},
	volume = {53},
	issn = {0025-5564},
	url = {https://www.sciencedirect.com/science/article/pii/0025556481900432},
	doi = {10.1016/0025-5564(81)90043-2},
	language = {en},
	number = {1},
	urldate = {2021-03-16},
	journal = {Mathematical Biosciences},
	author = {Robinson, D. F. and Foulds, L. R.},
	month = feb,
	year = {1981},
	pages = {131--147},
}

@article{gao_biological_2025,
	title = {Biological {Causes} and {Impacts} of {Rugged} {Tree} {Landscapes} in {Phylodynamic} {Inference}},
	url = {https://www.biorxiv.org/content/10.1101/2025.06.10.657742.abstract},
	doi = {10.1101/2025.06.10.657742},
	urldate = {2025-11-25},
	journal = {bioRxiv},
	publisher = {Cold Spring Harbor Laboratory},
	author = {Gao, Jiansi and Brusselmans, Marius and Carvalho, Luiz M. and Suchard, Marc A. and Baele, Guy and Matsen IV, Frederick A.},
	year = {2025},
}

@article{st_john_review_2017,
	title = {Review {Paper}: {The} {Shape} of {Phylogenetic} {Treespace}},
	volume = {66},
	issn = {1063-5157},
	shorttitle = {Review {Paper}},
	url = {https://doi.org/10.1093/sysbio/syw025},
	doi = {10.1093/sysbio/syw025},
	number = {1},
	urldate = {2025-12-11},
	journal = {Systematic Biology},
	author = {St. John, Katherine},
	month = jan,
	year = {2017},
	pages = {e83--e94},
}

@inproceedings{bocker_generalized_2013,
	address = {Berlin, Heidelberg},
	title = {The {Generalized} {Robinson}-{Foulds} {Metric}},
	isbn = {978-3-642-40453-5},
	doi = {10.1007/978-3-642-40453-5_13},
	language = {en},
	booktitle = {Algorithms in {Bioinformatics}},
	publisher = {Springer},
	author = {Böcker, Sebastian and Canzar, Stefan and Klau, Gunnar W.},
	editor = {Darling, Aaron and Stoye, Jens},
	year = {2013},
	pages = {156--169},
}

@article{brusselmans_importance_2024,
	title = {On the {Importance} of {Assessing} {Topological} {Convergence} in {Bayesian} {Phylogenetic} {Inference}},
	volume = {10},
	issn = {2057-1577},
	url = {https://doi.org/10.1093/ve/veae081},
	doi = {10.1093/ve/veae081},
	number = {1},
	urldate = {2025-12-22},
	journal = {Virus Evolution},
	author = {Brusselmans, Marius and Carvalho, Luiz Max and L. Hong, Samuel and Gao, Jiansi and Matsen IV, Frederick A and Rambaut, Andrew and Lemey, Philippe and Suchard, Marc A and Dudas, Gytis and Baele, Guy},
	month = oct,
	year = {2024},
	pages = {veae081},
}

@article{bordewich_computing_2007,
	title = {Computing the {Minimum} {Number} of {Hybridization} {Events} for a {Consistent} {Evolutionary} {History}},
	volume = {155},
	issn = {0166-218X},
	url = {https://www.sciencedirect.com/science/article/pii/S0166218X06003957},
	doi = {10.1016/j.dam.2006.08.008},
	number = {8},
	urldate = {2026-02-27},
	journal = {Discrete Applied Mathematics},
	author = {Bordewich, Magnus and Semple, Charles},
	month = apr,
	year = {2007},
	pages = {914--928},
}

@article{day_optimal_1985,
	title = {Optimal {Algorithms} for {Comparing} {Trees} with {Labeled} {Leaves}},
	volume = {2},
	issn = {1432-1343},
	url = {https://doi.org/10.1007/BF01908061},
	doi = {10.1007/BF01908061},
	language = {en},
	number = {1},
	urldate = {2026-02-27},
	journal = {Journal of Classification},
	author = {Day, William H. E.},
	month = dec,
	year = {1985},
	pages = {7--28},
}

@article{bryant_splits_2004,
	title = {The {Splits} in the {Neighborhood} of a {Tree}},
	volume = {8},
	issn = {0219-3094},
	url = {https://doi.org/10.1007/s00026-004-0200-z},
	doi = {10.1007/s00026-004-0200-z},
	language = {en},
	number = {1},
	urldate = {2026-02-27},
	journal = {Annals of Combinatorics},
	author = {Bryant, David},
	month = may,
	year = {2004},
	pages = {1--11},
}

@article{collienne_ranked_2024,
	title = {Ranked {Subtree} {Prune} and {Regraft}},
	volume = {86},
	copyright = {All rights reserved},
	issn = {1522-9602},
	url = {https://doi.org/10.1007/s11538-023-01244-2},
	doi = {10.1007/s11538-023-01244-2},
	language = {en},
	number = {3},
	urldate = {2026-02-24},
	journal = {Bulletin of Mathematical Biology},
	author = {Collienne, Lena and Whidden, Chris and Gavryushkin, Alex},
	month = jan,
	year = {2024},
	pages = {24},
}

@book{steel_phylogeny_2016,
	address = {Philadelphia, PA},
	title = {Phylogeny: {Discrete} and {Random} {Processes} in {Evolution}},
	isbn = {978-1-61197-447-8 978-1-61197-448-5},
	shorttitle = {Phylogeny},
	url = {http://epubs.siam.org/doi/book/10.1137/1.9781611974485},
	doi = {10.1137/1.9781611974485},
	language = {en},
	urldate = {2026-08-19},
	publisher = {Society for Industrial and Applied Mathematics},
	author = {Steel, Mike},
	month = sep,
	year = {2016},
}

@article{de_jong_neighborhoods_2016,
	title = {Neighborhoods of {Phylogenetic} {Trees}: {Exact} and {Asymptotic} {Counts}},
	volume = {30},
	issn = {0895-4801, 1095-7146},
	shorttitle = {Neighborhoods of {Phylogenetic} {Trees}},
	url = {http://epubs.siam.org/doi/10.1137/15M1035070},
	doi = {10.1137/15M1035070},
	language = {en},
	number = {4},
	urldate = {2026-08-19},
	journal = {SIAM Journal on Discrete Mathematics},
	author = {De Jong, J. V. and McLeod, J. C. and Steel, M.},
	month = jan,
	year = {2016},
	pages = {2265--2287},
}

\end{document}